# Smart Cities, Smart Libraries and Smart Knowledge Managers: Ushering in the neo-Knowledge Society

Mayukh Bagchi

Institute for Globally Distributed Open Research and Education *(IGDORE)*

[mbagchi.research@gmail.com](mailto:mbagchi.research@gmail.com)

[mayukh.bagchi@unitn.it](mailto:mayukh.bagchi@unitn.it)

**Abstract:**

The emergence of smart cities as a specific concept is not very old. In simple terms, it refers to cities which are sustainable and driven predominantly by their Information and Communication Technology (ICT) infrastructure. Smart libraries and smart knowledge managers, alongside its other smart component-entities, are vital for their emergence, sustenance and progress. The paper attempts at deducing a symbiosis amongst smart cities, smart libraries and smart knowledge managers. It further elaborates on how these will usher in the neo-knowledge society, and the opportunities it'll offer vis-à-vis Library and Information Science (LIS). Finally, it concludes on an optimistic note, mentioning possible future research activities in this regard.

1. **Introduction:**

Cities have remained the epicentre of human life and action since the dawn of human civilization. With passing time and improving social intellect, they have been driven through and viewed from a variety of perspectives. One such prominent view-point remains the infallible conceptual hierarchy of Data-Information-Knowledge-Wisdom (DIKW) (Ackoff 1989). The earliest societies were the very first transactors of data in their most rudimentary form- facts, figures or instructions. Such a trend continued for a long time until the early and late medieval societies realised the importance of contexts and relationships amongst the data, with the focus shifting to contextualized and meaningful data- Information. As societies gradually modernized, aided by industrial and technological innovation (especially, computing prowess), information as the sole basis of communication and transaction, turned out to be functionally inefficient. Knowledge, i.e. experience, context, intuition and insight applied to information, became the dominating driver of the society. Corporations and public institutions invested heavily in researching and developing the various possible tools and technologies aimed at mining out knowledge from data and information, leading to the emergence of new realms of study like Knowledge Discovery and Data Mining. Thus, the natural progress of societies complemented and correlated with the progression envisioned in the DIKW conceptual hierarchy.

That's not the end, though. The arrival of the 21st century ushered in multi-dimensional changes to the fabric of societies. Data, Information and Knowledge became equally important tenets around which societies revolved, and the concept of Smart Cities came into existence. Data once more became the talk of societies due to its unbridled generation and capture, facilitated by new technological concepts like Big Data, Internet-of-Things (IoT) and the highly user-participative social media boom. A brand new discipline of Data Science came up to support its research and development. The demand for information and knowledge became even more stronger, considering their emerging economic importance as resources and commodities. Libraries and librarians, traditionally limited by their role as preservers and disseminators of society's recorded knowledge, donned the hats of Knowledge Resource Centres, Data Librarians, Information Scientists and Knowledge Managers, dabbling in the life-cycle management of unimaginably diverse forms of information and knowledge. Even the concept of wisdom, which is understood as collaboratively applied knowledge enriched by experience, foresight and heuristics, has been latently leveraged in the form of predictive analytics. The mix and match of Information and Communication Technology (ICT), sustainability, innovation economy, good governance and Quality-of-Life (QoL) metrics contributed towards a newer variant of knowledge-based society, a neo-knowledge society. In a nutshell, social life became way more smarter and technologically rich. The paper attempts to present a cohesive yet multi-faceted view of how smart libraries and smart knowledge managers can foster smart cities into ushering in the neo-knowledge society.

## 2. Review of literature:

There have been a considerable number of works of broad conceptual importance, dwelling upon various dimensions of smart cities, smart libraries, knowledge managers and knowledge societies, individually. Though, very few have attempted to properly delineate how the interactions (both intra and inter) amongst these dimensions and concepts, ultimately lead to the emergence of the neo-knowledge society.

Eremia, Toma and Sanduleac (2017) discuss the emergence and the best-fitting characteristics of smart cities through an evolutionary analysis. The US-India Business Council (USIBC), in a white paper, defines several unique markers characteristic of a smart city, and goes on to discuss the important role of ICT and smart intellectual infrastructure in envisaging them, with a special reference to India. An all-encompassing reference framework for conceptualizing smart cities in India has been proposed in the landmark report "Reconceptualising Smart Cities: A Reference Framework for India" (Bhattacharya *et al.* 2015). An overview of the various challenges and associated assessment metrics for smart cities have been illustrated in Monzon (2015). Novotný, Kuchta and Kadlec (2014) address aspects related to technological solution, support and applications required for smart cities. Societal, technological and market requirements to analyse, assess and formulate new standards for smart cities have also been documented (ISO/IEC JTC 1,

2015). TU Wien, University of Ljubljana and TU Delft (2007) in the report of one of their leading research projects, ranked several smart cities geographically distributed across Europe, based on selected parameters.

Leicestershire County Council, in one of their library and information service reports, discuss the nature, characteristic features and working of an ideal smart library. Schöpfel (2018) attempts at a mapping of the essential dimensions of smart cities and smart libraries, enriched through an integration of aspects from people, processes and technologies. Indergaard (2012) elucidate several ideas and strategies for a smooth transition to a new library concept suited for smart cities. Breeding (2017) emphasize the role of technologies in achieving smart libraries, alongside the need to develop efficient technical mechanisms for library data privacy in the smart era. Yang *et al.* (2017) identifies the various possible applications of Deep Learning in smart libraries. Baryshev (2015) elaborate, in a detailed manner, the developmental journey from electronic libraries to smart libraries. Evers (2002) charts out the natural growth of a knowledge society and the ensuing knowledge gap, from a combination of economic and epistemological perspectives. The Department of Social and Economic Affairs of the United Nations (2005), in a landmark monograph, explains the changing developments affecting the conceptual length and breadth of knowledge, and how they will accentuate the development of futuristic knowledge societies. The concept of knowledge societies have also been attributed to the emergence of the concept of knowledge-based economies, their challenges, best practices and success models (Garner 2016). Wessels *et al. (*2017) explore the interrelationships between open data and big data within a generic data ecosystem, facilitating the emergence of knowledge societies. Hornidge (2011) attempts a rich historical and conceptual review of knowledge societies from an academic perspective.

### 3. Smart Cities, Smart Libraries & Smart Knowledge Managers - The Symbiosis:

*3.1 Smart Cities:*

The concept of smart cities, since its popular inception in the 1990s, has been characterized and interpreted differently by different stakeholders. Batty *et al.* (2012) define a smart city as one which is characterized by the embedding of ICT within conventional city infrastructures, focussing on the integration and potentiality of applicable digital technologies. The American Planning Association (APA), in several of its regional conferences, have also visualized smart cities as an interactive dual of cities- with its citizens, services and sub-systems, aided by an integrated ICT infrastructure as its backbone. The much discussed IBM Smart Cities documentation speaks of a smart city as a knowledge-optimized liveable infrastructure, with efficient analytics-support and information management-based governance to drive the city's subsystems. NASSCOM ("National Association of Software and Service Companies"), India, have also gone to the extent of developing an integrated ICT and geospatial technology framework, specially designed for the Indian smart cities programme. Technological utilities .which are poised to be major players in a smart city environment include (not exhaustive) Artificial Intelligence (AI), Big Data Analytics, Blockchains, Internet-of-Things (IoT), Cloud

Computing, Knowledge Management (KM) tools and techniques, Semantic Web technologies, Agent technologies, Robotics, Networking technologies, Ubiquitous Computing, Geo-Spatial Information Systems and Human Computer Interaction (HCI).

Besides the stress on ICT and upcoming digital technologies as the core building-blocks, explanations characterizing smart cities have also included many other aspects, some of which are social infrastructure, environmental sustainability, smart policy making, energy efficiency, mobility and equity optimization, and smart economy.

3.2 *Smart Libraries:*

Smart libraries are libraries envisioned for smart cities. They can be aptly defined using a re-formulation of Prof. S.R. Ranganathan's library triad - smart information, smart users and smart staff. As an overall concept, smart libraries are considered as public information institutions piloting the accession, dissemination and curation of data, information and knowledge (mostly in electronic form) for users who are technology friendly, facilitated by a combination of advanced ICT tools and technologically sound library staff, acting as a centre of thrust and support for the sustainability of smart cities. Some of the distinctive as well as evaluative hallmarks of smart libraries, which are in stark contrast to the reality of majority of today's libraries are enumerated as follows:

- All-encompassing and pervasive use of ICT, resulting in intelligent and automated library housekeeping operations
- As the central hub of a city's technology and innovation management, providing unprecedented level of open access to information through open science cloud(s) & smart digital repositories
- As the backbone of a smart city's Intellectual Infrastructure, piloting of increasingly important user-centric smart practices such as Information & Knowledge management- including services like Digital Curation, Research Data Management (RDM) etc.
- As a smart city's go-to centre for learning, skill, curriculum and research development, facilitated by in-house developed & externally aggregated formal, information and virtual open educational packages in the form of Massive Open Online Courses (MOOCs), Open Educational Resources (OERs), educative mashups and Information, Media and Data literacy programmes
- Community engagement practices through smart community information services, leveraging technologies like semantic web, linked open data and HCI
- Architecturally and indoor décor-wise sustainable, truly implementing the essence envisioned in conceptualizations like makerspaces
- Smart knowledge managers from the library and information science community with expertise in applying advanced technologies, spirited by the motto of providing smart information services as a part of good, inclusive & sustainable governance.

3.3 *Smart Knowledge Managers:*

Smart libraries would require smart knowledge managers- librarians well versed with emerging techno-managerial skill sets and innovative learning pedagogies, over and above their traditional core competencies. In manning libraries driven by ICTs, the profile of smart knowledge managers would include tasks like formulation & implementation of knowledge-based programmes, anticipative & adaptive innovation-driven knowledge strategies, and self-learning & problem-solving worthy knowledge communication. There have been several analyses regarding the competency-spectrum of smart knowledge managers. One such study (Al-Hawamdeh and Foo 2001) develops two relevant skill-set formulations. The linear formulation enumerates skills like creativity, ICT, diverse subject backgrounds, analysis and communication to be of utmost essentiality for smart knowledge professionals. The second approach, which is more taxonomic in nature, divides the professional competencies into six broad divisions- information skills, communication skills, analytical skills, IT skills, leadership-management abilities and personal characteristics- which are further sub-divided to form a more granular view-point. Further, in mapping with the characteristic hallmarks of smart libraries, smart knowledge managers are also expected to be proficient in data and information governance, knowledge management, web technologies (including semantic web), artificial intelligence, data analytics, e-governance & e-learning technologies, digital curation, digital rights management, open science, makerspaces, and, information & media literacy. It is also noteworthy that the essence of the emergence of smart cities also calls for decentralization within libraries, with commissioning of specialized posts like Data Librarian, Metadata Librarian, Information Literacy Librarian, Technology Librarian, Librarian for specially abled individuals- each devoted to an area of expertise within a library, as opposed to the mostly linear organizational structure in majority of today's libraries.

*3.4 The Symbiosis:*

With the continual rise in excitement regarding smart cities across various quarters, it is but natural that discussions about smart libraries and their custodians are also happening, albeit mostly within the library and information science community. The transformation of cities into smart cities will also bring about metamorphoses in today's libraries, basing them on a strong technological foundation, becoming more user-centric and fulfilling a city's information needs as its unshakeable intellectual pillar. Smart Libraries and smart knowledge managers also, as vital components of a smart city, should support the smooth running of its multi-dimensional services facilitating sustainability, inclusiveness and good governance for all its citizens.

## 4. The neo-Knowledge Society:

A knowledge society, as popularly understood, refers to a society which thrives by leveraging knowledge through an intricate succession and implementation of knowledge management activities like knowledge creation, knowledge capture, knowledge codification and retention, knowledge sharing and knowledge use. It is also often re-interpreted as a society which deals with knowledge, both as resource and as commodity, having different societal ramifications with respect to technology, business, politics, economics, culture and education, with an ultimate aim of accentuating human progress. Such a society is different from an information society, in the sense that the latter posits creation, integration, transformation, management, distribution, use and curation of information (i.e. meaningful and organized data) as its prime driver, whereas the former deals in uncovering & putting-to-use contextualized and experience-laced information, i.e. knowledge, one level up in the abstraction hierarchy (DIKW). Data Mining- finding out patterns indicative of knowledge in a huge corpora of information- remains the pre-eminent technology in knowledge societies.

The neo-knowledge societies, which smart cities are expected to usher in, retains the characteristics of the present conceptualization around knowledge societies, and ventures further in terms of its few extra transformational characteristics. One of its envisioned characteristic features lays out the equal importance of data, information and knowledge in the conduct and management of various activities in different walks of societies. A smart city environment would invariably be witness to an unprecedented gamut of data generation, capture, analysis and curation due to enhanced understanding, access and implementation of advanced technological arenas like Data Science, Big Data Analytics and IoT. Advances in computing-based information and knowledge processing would similarly induce a paradigm shift in governing the complex network of information flow expected in a smart city. In addition to these features, another hallmark of the neo-knowledge society would also be the potential social involvement of the concept of wisdom, in diverse forms like predictive analytics, predictive modelling and business intelligence; to extract insights from knowledge and providing decision support services for the smart cities to remain competitive as well as relevant. It is needless to mention that to make all these happen, smart libraries and smart knowledge managers have a pivotal role to play in terms of supporting informed decision-making and soft-power development for neo-knowledge societies.

## 5. Opportunities vis-à-vis Library and Information Science (LIS) as a domain:

The ushering in of a true neo-knowledge society in the form of smart cities, supported by smart libraries would also signal an opening to a plethora of professional and research opportunities for Library and Information Science as a domain. Some of the relevant aspects are categorized and discussed in the following points:

*5.1 LIS Schools & Curricula:*

First of all, to achieve smartness for libraries in technological terms, LIS schools and their curricula should undergo major revamp. Old fashioned syllabi and teaching methodologies should undergo continuous revision, and, make way for topics (essentially, smart library technologies mentioned in the previous sections) and learning pedagogies with a technological base, in sync with the needs of smart libraries. This does not necessarily translate into shelving all traditional arenas of study in LIS, rather, it implies technological conditioning of selected topics in order to make them relevant and applicable, alongside new introductions. Topics which are trending yet difficult to accommodate within the formal teaching curricula can be taken up and discussed in academic seminars and colloquiums. There is also an emergent need for LIS schools to re-brand and market themselves for increased visibility, exposure and funding (case in point, i-schools). Further, intra-disciplinary and interdisciplinary conclaves and conferences should be organized by these schools from time-to-time, with an aim to motivate its students to become future smart knowledge managers.

*5.2 LIS Research:*

Similar to goals envisioned for LIS schools & curricula, research conducted in Library and Information Science should also orient itself towards tackling libraries of the future, intelligently balancing the delicate line between concept build-up and practically implementable models. Researchers should immerse themselves into how upcoming technologies, frameworks, standards and concepts can be leveraged by libraries, making them technologically rich enough to be able to match the growing standards of smart cities, and pitch for relevant projects showcasing their competencies. There should also be serious research into how traditional LIS concepts can be mapped and made relevant with emerging technologies. Further, it will be a best case scenario for all stakeholders if research and curriculum-based education go hand-in-hand in an LIS school.

*5.3 LIS Job Opportunities:*

It is beyond doubt that smart cities will bring with it unbridled employment opportunities in multiple sectors for professionals skilled in Library and Information Science. Besides their obvious chances in smart libraries & academia, graduates and researchers schooled in updated curricula and research trends are also expected to join IT consultancies, policy think tanks, scientific institutions, knowledge management divisions of corporate bodies, and, even launch their own start-ups, providing consultancy services to libraries and allied institutions.

## 6. Conclusion and Future Work:

Before concluding, it would be relevant to note that the transformation of today's cities towards their smart future is inevitable, and that the role of smart libraries and smart knowledge managers will be paramount in that set-up. Hence, active future research is needed in the form of implementable frameworks as to how the emerging technologies can be integrated and leveraged to deliver smart solutions in every facet of library services.

Wessels et al. (2017). Open Data and the Knowledge Society. Amsterdam University Press.

Yang, X. et al. (2017). Smart Library: Identifying Books on Library Shelves Using Supervised Deep Learning for Scene Text Reading. 2017 ACM/IEEE Joint Conference on Digital Libraries (JCDL). doi:10.1109/jcdl.2017.7991581

**Further Readings:**

Bagchi, M. (2020). Conceptualising a Library Chatbot using Open Source Conversational Artificial Intelligence. DESIDOC Journal of Library & Information Technology, 40(6).

Bagchi, M. (2019). A Knowledge Architecture using Knowledge Graphs. MS Dissertation, Indian Statistical Institute.

Bagchi, M. (2021). A Large Scale, Knowledge Intensive Domain Development Methodology. Knowledge Organization, 48(1), 8-23.

Bagchi, M. (2021). Towards Knowledge Organization Ecosystem (KOE). Cataloging & Classification Quarterly, 1-17.

Satija, M. P., Bagchi, M., & Martínez-Ávila, D. (2020). Metadata management and application. LIBRARY HERALD, 58(4), 84-107.

Bagchi, M. (2022). Open Science for an Open Future. In Proceedings of the International Conference on Exploring the Horizons of Library and Information Sciences: From Libraries to Knowledge Hub (SRR125) (pp. 422-431). EasyChair Preprint: 7351